\documentclass[11pt]{article}
\usepackage{amsmath,amsthm}
\usepackage{latexsym}
\usepackage{amssymb}
\usepackage{array}
\usepackage{epsfig}

\newtheorem{theorem}{Theorem}
\newtheorem{lemma}{Lemma}[theorem]

\newcommand{\Epsfig}[5] {
                          \begin{figure}[#2]
                              \epsfxsize=#5
                              \centerline{\epsfbox{#1}}
                              \caption{\bf  #3 \label{#4}}
                          \end{figure}
                        }

\newcommand{\Sec}[2]{ \section{#1} \label{#2}}
\newcommand{\eq}[1]{(\ref{#1})}
\newcommand{\Th}[1]{theorem (\ref{#1})}
\newcommand{\makev}[2]{\left(\begin{array}{c} #1 \\ #2 \end{array}\right)}
\newcommand{\hen}{H\'{e}non }

\def \bdz{{\mathbf{\zeta}}}

\def \bs {{\mathbf{s}}}
\def \bT {{\mathbf{T}}}

\def \bx {{\mathbf{x}}}

\def \bz {{\mathbf{z}}}

\def \R {{\mathbb R}}

\def \Z {{\mathbb Z}}

\def \S {{\mathbb S}}
\def \qed {$\Box$ \\ \\}

\begin{document}

\title{Computing periodic orbits using the anti-integrable limit}
\author{D. Sterling \& J.D. Meiss
\thanks{Useful conversations with Robert Easton, Bruce Peckham and Holger 
Dullin are gratefully acknowledged. JDM was supported in part by NSF grant
number DMS-9623216.}\\ 
Department of Applied Mathematics\\
University of Colorado\\
Boulder, CO 80309
}
\date{February 6, 1998}
\maketitle

\begin{abstract}
Chaotic dynamics can be effectively studied by continuation from an 
anti-integrable limit.  Using the \hen map as an example, we obtain a 
simple analytical bound on the domain of existence of the horseshoe 
that is equivalent to the well-known bound of Devaney and Nitecki.  We 
also reformulate the popular method for finding periodic orbits 
introduced by Biham and Wenzel.  Near an anti-integrable limit, we 
show that this method is guaranteed to converge.  This formulation 
puts the choice of symbolic dynamics, required for the algorithm, on a 
firm foundation.\\
\textbf{AMS classification scheme numbers}: 58F05, 58F03, 58C15
%
%
\end{abstract}

%
\Sec{Introduction}{introsec}

The anti-integrable (AI) limit of a mapping is a singular limit in which 
the dynamics is not deterministic\cite{Aubry95}.  Each orbit in the AI 
limit is given by a sequence of symbols, and the dynamics becomes the 
unrestricted shift operator on the symbols.  It is well known that 
when the AI limit is ``nondegenerate'' many of the AI states can be 
continued away from the limit, becoming orbits of the original mapping 
\cite{Aubry90, MacKay92}. This continuation theory, a 
consequence of the implicit function theorem, is much simpler 
than the corresponding continuation from an ``integrable limit'', 
which often requires the machinery of KAM theory.

Biham and Wenzel \cite{Biham89} introduced a technique for finding 
periodic orbits of maps, in particular maps that are obtained from a 
variational principle.  They generalized the notion of ``gradient 
search'' to find the minimum of the variational function by 
introducing a ``pseudo-gradient'' system of differential equations.  This 
is obtained by multiplying the gradient by a diagonal matrix of signs.  
The basic idea is that this will allow one to find critical points 
other than the minimum.  This method has become popular 
\cite{Biham90,Davis91,Wenzel91,Biham92,Dey95,
DAlessandro90,Hansen92,Politi92,Hunt96,Skodje90,Fang94,Kaplan93,Nagai97}
in spite of the fact that it has no rigorous foundation.  Indeed, Grassberger et. 
al \cite{Grassberger89} found examples where the method 
fails to find unique orbits for certain parameter values in the \hen map.

In this paper we use the \hen map as an example to illustrate the 
usefulness of the AI limit.  At the AI limit, the \hen map reduces to 
the full shift on two symbols, and we study orbits that continue 
from these states.  We obtain an explicit bound on this continuation 
which happens to correspond exactly to the bound on the existence of 
the horseshoe obtained by Devaney and Nitecki \cite{Devaney79} who used a 
geometrical argument.  We then reformulate the Biham and Wenzel 
method and show that the pseudo-gradient system has a hyperbolic sink 
at the AI limit, and that this sink persists at least for the same parameter 
range for which we can continue the orbits.

%
\Sec{Anti-Integrable Limit}{aisec}

The anti-integrable (AI) limit can be formulated in two ways.  
Consider a map, $f: M \rightarrow M$ on a $d$ dimensional manifold $M$ 
that depends upon some parameters.  The general idea is to rewrite the 
system as an equivalent implicit relation $F: M\times M \rightarrow 
\R^{d}$ in such a way that $F$ becomes singular when a parameter, 
say $\epsilon$, is zero \cite{Aubry95}.  For example if $F(x,x') = \epsilon 
G(x,x') + H(x)$, then the ``orbit'' at $\epsilon=0$ corresponds to any 
sequence of zeros of $H$--the dynamics is not deterministic.  We say 
that $\epsilon=0$ corresponds to the {\it anti-integrable limit} of 
the map $f$.  If the derivative of $H$ is nonsingular, then a 
straightforward implicit function argument can be used to show that 
(at least some of) the AI orbits can be continued for $\epsilon \neq 
0$ to orbits of the map $f$ \cite{Aubry91, MacKay92}.  An AI limit 
with this property is called {\it nondegenerate}.

Maps that are derived from a Lagrangian variational principle often 
have an AI limit.  For example, orbits of the \hen map, written in the 
form
\begin{equation}\label{henonmap}
   \makev {{x'}}{{y'}} = \makev{ {y-k+x^2} }{ {-bx}} \,,
\end{equation}
can be obtained as the critical point of an action,
\begin{equation}\label{action}
    W[\bx] = \sum_{t} b^{-t} L(x_{t},x_{t+1})    \,,
\end{equation}
where we label time along the trajectory by a subscript $t$, and the 
factor $b^{-t}$ allows the system to be non-area preserving.  An orbit 
$\bx = \{\ldots x_{t},x_{t+1},\ldots\}$ is a critical point of $W$, 
and $y_{t} = bL_2(x_{t-1},x_t) = -L_1(x_t,x_{t+1})$.  
When the Lagrangian $L$ can be put into the form $L(x,x') = \epsilon 
T(x,x') - V(x)$, then $\epsilon=0$ corresponds to an AI limit 
\cite{MacKay92}.  This form is quite natural for ``mechanical 
systems'' where $T$ represents kinetic and $V$ potential energy.

For example, the Lagrangian for the \hen map is
\begin{eqnarray}\label{lagrangian}
    \hat L(x,x') &=& -x x' + \frac13 x^{3} -kx  \,.
\end{eqnarray}
Critical points of the action give the 
second order difference or Euler-Lagrange form of the map.
To put \eq{lagrangian} into a form appropriate for the AI limit, we 
rescale, defining
\[
       z=\epsilon x \quad,\quad \epsilon=\frac{1}{\sqrt{k}} \,,
\]
to obtain the new Lagrangian,
\begin{eqnarray}\label{scaledlag}
    L(z,z') = \epsilon^{3}\hat{L}(x,x') = -\epsilon z z' + \frac{z^3}{3} - z \,.
\end{eqnarray}
Critical points of the action correspond to sequences $z_t$ such that
\begin{equation}\label{scaledmap}
   -\epsilon (z_{t+1} +b z_{t-1}) + z_t^2 -1  = 0 \,.
\end{equation}
Of course this implicit system is easily solved for $z_{t+1}$ whenever 
$\epsilon \ne 0$ to give an explicit dynamical system.  However, when 
$\epsilon = 0$ \eq{scaledmap} no longer defines a map, even though it 
still represents a critical point of $W$.  This is the AI limit of 
the \hen map.  At this limit, ``orbits'' correspond to any bi-infinite 
sequence $\bz \in \S$, where we denote the set of bi-infinite 
sequences with two symbols by
\[
\S = \{ \bs: s_t\in\{1,-1\} \,, t \in \Z \} \,.
\]

Each $\bs \in \S$ continues to a unique 
orbit $\bz(\epsilon)$ where $\bz(0) = \bs$ when $\epsilon$ is small 
enough \cite{Aubry95, MacKay92}.  In fact, we can obtain an explicit upper 
bound on the range of $\epsilon$ for which this correspondence is 
guaranteed. It will be appropriate to use the $l^{\infty}$ norm, $||\bx||_{\infty} 
= \sup_{t}|x_{t}|$, and define $B_{M}$ to be the closed ball of radius 
$M$ around the point $\bs$,
\begin{equation}\label{BMdef}
    B_{M}(\bs) = \{{\bz}  : ||\bz - \bs||_{\infty} \le M \} \,.
\end{equation}
With this notation, we can prove:

%
%
\begin{theorem}
\label{existtheorem}
For every symbol sequence $\bs \in \S$, there exists a corresponding 
unique orbit, $\bz(\epsilon)$, of the \hen map \eq{scaledmap} such 
that $\bz(0) = \bs$ providing
\begin{equation}\label{epsmax}
     |\epsilon|(1+|b|) < \gamma_{\infty} \equiv 2\sqrt{1-2/\sqrt{5}} \approx 0.649839  \,.
\end{equation} 
The orbit $\bz(\epsilon)$ is contained in the ball $B_{M_{\infty}}(\bs)$ where
\begin{eqnarray}\label{ballsize}
      M_{\infty} & = & 1 - \sqrt{1 -\gamma\frac{\gamma+\sqrt{\gamma^2+4}}{2}} \,,\\
       \gamma    & \equiv & |\epsilon|(1 + |b|) \,.
\end{eqnarray}
\end{theorem}

\noindent
To prove the theorem, we will write orbits $\bz(\epsilon)$ of the \hen map as
fixed points of an operator $\bT$ whose $t^{th}$ component is 
\begin{equation}\label{Tdef}
  T_t(\bz) \equiv s_t\sqrt{1 + \epsilon(z_{t+1} + bz_{t-1})} \,,
\end{equation}
where we choose the sign of the square root using a sequence $\bs \in 
\S$.  When $\epsilon = 0$, the operator becomes $\bT(z) = \bs$, which 
trivially has a unique fixed point corresponding to the AI state, $\bz 
= \bs$.  We will use the contraction mapping theorem to show that for 
small enough $\epsilon$ the fixed point persists.  The first step in 
the proof is to show that there is a domain ${\cal C}_1$ of the parameters
$(\gamma,M)$ such that $\bT$ is a contraction on $B_{M}(\bs)$; this domain 
is illustrated in Fig.  \ref{contbound}.  This is improved by bounding 
the domain ${\cal C}_n$ in the $(\gamma,M)$ plane where $\bT^n(\bz)$ 
is a contraction.  It is easy to see that
\[
 {\cal C}_n \subseteq {\cal C}_{2n} \,;
\]
thus we can improve the bound on $\gamma(\epsilon)$ by iteration of $\bT$.  
Figure \ref{contbound} illustrates the parameter domains that we find 
for $n=1,2,30$, and $\infty$. The maximal $\gamma$ values for each of
these iterates are denoted $\gamma_{n}$ in Fig. \ref{contbound}.

\Epsfig{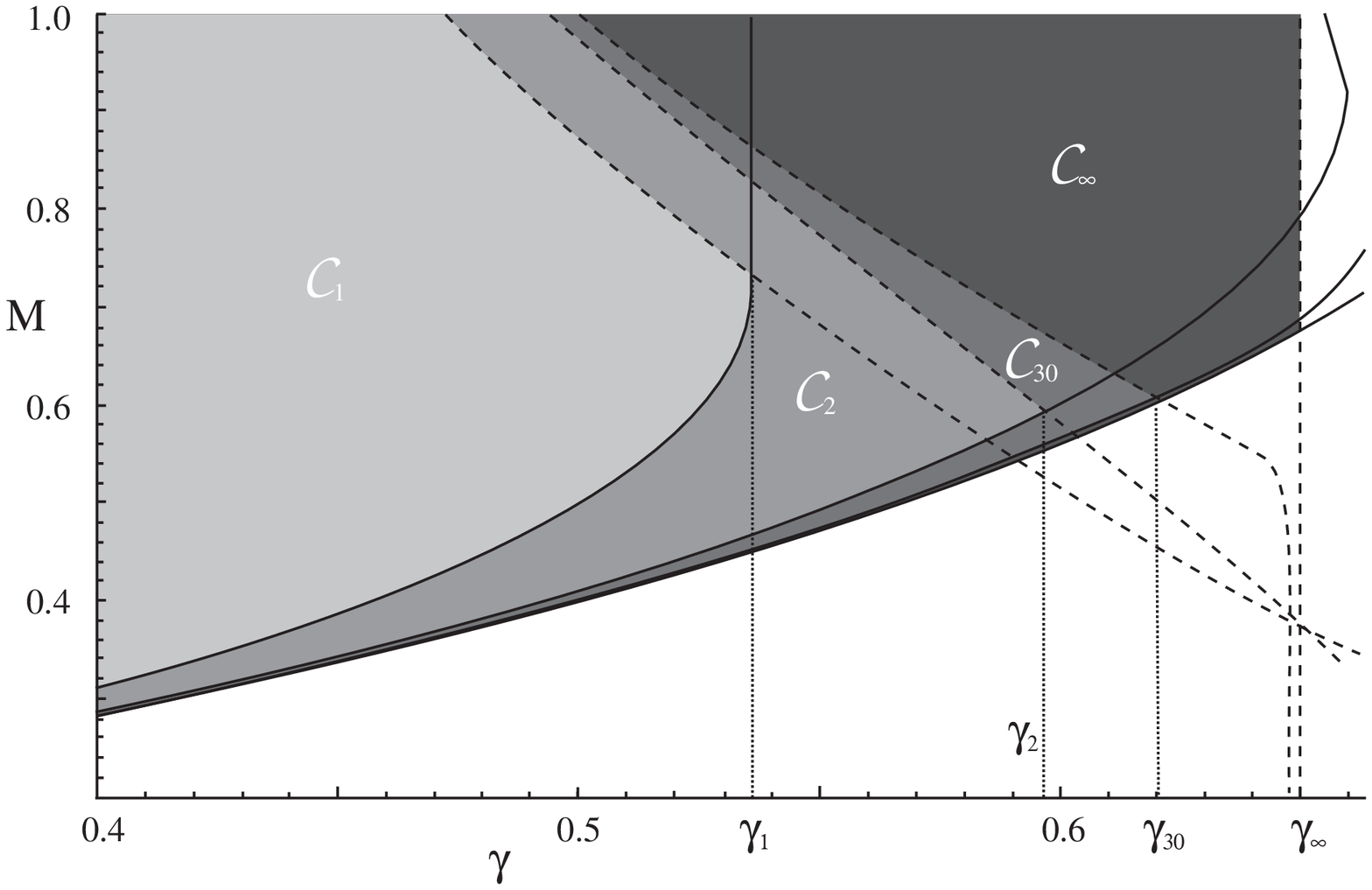}{ht}
{Contraction regions ${\cal C}_{n}$ for $b=1$ and $n=1,2,30,\infty$.}
{contbound}{4.5in}

To demonstrate that $\bT^n$ is a contraction mapping we must show that 
for $(\gamma,M) \in {\cal C}_n$
\begin{enumerate}
\item $ \bT^n: B_{M}(\bs) \rightarrow B_{M}(\bs)$
\item $ ||D\bT^n(\bz)||_\infty < 1 $ for all $\bz \in B_{M}(\bs)$
\end{enumerate}
The first requirement gives the solid curve in the 
figure defining a lower bound on $M$ values in ${\cal C}_n$ as well 
as right boundary where the range of $\bT^{n}$ is no longer real. 
The dashed boundary of ${\cal C}_n$ is given by the second requirement.
The maximal value of $\gamma$ is obtained in the limit $n 
\rightarrow \infty$.

We begin with a lemma to prove the first requirement.
%
%
\begin{lemma}
\label{Tboundlemma}
For any $|\gamma|< 1/\sqrt{2}$, and any $M > M_{\infty}(\gamma)$ there is an $N$ 
such that $\bT^n: B_{M}(\bs) \rightarrow B_{M}(\bs) \, , \,\forall n>N$.
\end{lemma}
\noindent
\underline{Proof:}

Let $\bT$ be defined by \eq{Tdef}. For any $\bz \in B_M(\bs)$  it is 
easy to see that
\begin{equation}\label{tbounds}
   \alpha_{n} \le ||\bT^{n}(\bz)||_{\infty} \le \beta_{n}  \,,
\end{equation}
where the sequences $\alpha_{n}$ and $\beta_{n}$ are given by the 
iterations
\begin{eqnarray*}
   \beta_{n+1}   &=& f(\beta_n) \equiv \sqrt{1 + \gamma \beta_n}  \,,\\
   \alpha_{n+1}  &=& \sqrt{1-\gamma \beta_{n+1}}  \,,
\end{eqnarray*}
with the initial conditions  $\beta_0=1+M$ and $\alpha_{0}=1-M$.
The map $f(\beta)$ has a single attracting fixed point
\[ 
     \beta_\infty = \frac{\gamma + \sqrt{\gamma^2 + 4}}{2}  \,.
\]

Each of the $\alpha_{n}$ must be real, so we must have $1-\gamma 
\beta_{n} \ge0$.  This requirement gives a right boundary to the region 
in the $(\gamma,M)$ plane where $\bT^{n}$ exists.  As $n \rightarrow 
\infty$ these boundaries approach the vertical line defined by
\begin{equation} \label{realbound}
      1 - \gamma \beta_\infty = 0 \Rightarrow \gamma = 1/\sqrt{2}  \,,
\end{equation}
which gives one of the bounds in the lemma.

Finally, \eq{tbounds} implies that
\[
      ||\bT^n - \bs||_\infty \le \max{(|\alpha_n-1|,|\beta_n-1|)} = 1-\alpha_n  \,.
\]
Thus the requirement that $\bT^{n}$ maps $B_{M}(\bs)$ into itself gives the 
implicit relation $1 - \alpha_n \le M$. As $n$ approaches infinity, 
these domains approach the region
\[
M \ge M_{\infty}(\gamma) = 1 - \alpha_\infty \,.
\]
For each $\gamma$ the sequence $M_{n}$ for which $1-\alpha_n-M_{n} 
= 0$ converges monotonically to $M_{\infty}$ from above; therefore 
for any $M > M_{\infty}(\gamma)$, there is an $N$ such that 
$\bT^n:B_M(\bs) \rightarrow B_M(\bs) \, , \, \forall n>N$.
\qed

%
%
\noindent
\underline{Proof of \Th{existtheorem}}
Let $B_M(\bs)$ be defined by \eq{BMdef} and $\bT$ by \eq{Tdef}.  The map 
$\bT^n$ is a contraction if $||D\bT^n(z)||_\infty < 1$ for all $\bz 
\in B_{M}(\bs)$.  Using the chain rule gives
\begin{equation} \label{chainrule}
    ||D\bT^n||_\infty \le \frac{\gamma^n}{2^n\prod^n_{j=1} \alpha_j} 
\end{equation}
From this and lemma \ref{Tboundlemma}, the map $\bT^n$ is a contraction 
in the region ${\cal C}_n$ given by 
\begin{equation}
\label{Cndef}
   {\cal C}_n \equiv 
    \{ (\gamma,M) : 
         1-\alpha_n \le M  \,,\,
         \gamma \le 2 \prod_{j=1}^{n} \alpha_j^{1/n} \,,\,
         \alpha_{j} \mbox{ Real for } j \le n \}  \,.
\end{equation}
We find the boundary of ${\cal C}_\infty$ by noting that the 
product in \eq{Cndef} converges to $\alpha_{\infty}$ because the 
sequence $\alpha_{j}$ converges geometrically to the fixed point.  
Thus there exists an $N$ such that the map $\bT^{n}$ is a contraction 
for all $n>N$ whenever
\begin{equation}\label{bound1}
    \gamma < 2 \alpha_{\infty} \,.
\end{equation}
Using the form for $\alpha_{\infty}$ we obtain the bound
\begin{equation}\label{gamBound}
   \gamma^{2} < 4(1-2/\sqrt{5}) \,,
\end{equation}
which directly gives \eq{epsmax}.  This bound is clearly more 
restrictive than \eq{realbound} so we can conclude that for any
\[ 
 (\gamma, M)\in {\cal C}_\infty = \{(\gamma,M): M > M_{\infty}(\gamma), \gamma < \gamma_{\infty} \}
\]
there is an $N$ such that $\forall n>N$, the map $\bT^{n}(\bz)$ is a 
contraction map whenever $\bz \in B_{M}(\bs)$ The Banach fixed point 
theorem then implies that $\bT^{n}$ has a unique fixed point in 
$B_{M_{\infty}}(\bs)$ for all $n > N$.  But if $\bz$ is a fixed point both 
of $\bT^{N+1}(\bz)$ and of $\bT^{N+2}(\bz)$ then it must be a fixed 
point of $\bT$.  Since this is true for an arbitrary symbol sequence, 
for any $\gamma$ in the range \eq{gamBound}, there is a corresponding 
unique orbit of the \hen map.
\qed

By a completely different argument, Devaney and 
Nitecki\cite{Devaney79} proved that the non-wandering set of the \hen 
map is hyperbolic and conjugate to the 2-shift for exactly the same 
range of $\gamma$ that we give in Theorem \ref{existtheorem}.

%
\Sec{The Method of Biham and Wenzel}{BW}

Biham and Wenzel \cite{Biham89, Biham90} introduced a popular 
algorithm for finding periodic orbits of maps. They present the method 
in four steps
\begin{enumerate}
        \item  Write the map in variational form with action $W[\bx]$. 
        Critical points of $W$ give the desired orbits.

        \item Use symmetries or some other technique to label the various 
        critical points of $W$ symbolically with a symbol sequence of 
        signs, $s_{j}$.

        \item  Choose an ``arbitrary'' initial guess for a periodic orbit, 
        $x_{j}(0), j=0,1,\ldots n-1$ of period $n$.
        
        \item  Introduce a pseudo-gradient dynamics with a fictitious 
        time $\tau$ to find an extremum of the action
        \begin{equation}\label{BWode}
           \frac{d\bx}{d\tau} =  -\mbox{Diag}(\bs) \nabla W[\bx] \,.
        \end{equation}
        For a period $n$ orbit, we set $x_{j+n}(\tau) = 
        x_{j}(\tau)$, and solve only the differential equations for $x_{0}, 
        x_{1},\ldots x_{n-1}$. 
\end{enumerate}

Note that when $\mbox{Diag}(\bs) = {\bf I}$ this method is a standard 
gradient search for finding a local minimum of $W$, and if we set 
$\mbox{Diag}(\bs) = -{\bf I}$, then we would simply be finding a local maximum of $W$. 
More generally, the symbols are chosen in an attempt to 
``identify the critical points'' of the action, but there is ``no general way 
to find'' the symbolic representation \cite{Biham90}.

The hope is that for each choice of $s_{j}$ (and some range of map 
parameter) the system \eq{BWode} either has a ``unique'' attracting 
fixed point, or else the ``corresponding'' periodic orbit does 
not exist.  This claim seems at least plausible for the \hen map, 
since it has at most $2^{n}$ fixed points of period $n$, and there are 
precisely $2^{n}$ choices of periodic sequences $\bs$.  However, 
Grassberger et.  al \cite{Grassberger89} have found examples for the 
\hen map where the method fails to find unique orbits at certain 
parameter values.  More generally, there is no guarantee
that the fixed point of \eq{BWode} corresponding to the orbit of 
interest is attracting. Moreover, this system of equations certainly 
can have other attractors, including chaotic ones.

A rigorous foundation for the Biham and Wenzel method can be given when the map has an AI limit.
Suppose the discrete Lagrangian has the form $L(x,x') = 
\epsilon T(x,x') - V(x)$, and that the potential $V$ has a 
discrete set of nondegenerate critical points $\mbox{Crit}(V) = \{c_0,c_{1},c_{2},\ldots 
c_{m}, c_{m+1}\}$, where we order the points, $c_{j}< c_{j+1}$ with the
convention that $c_0 = -\infty$ and $c_{m+1}= \infty$.
When $\epsilon = 0$ \eq{BWode} decouples, and becomes simply
\[
   \frac{dx_{j}}{dt} = - s_{j}V'(x_{j}) \,.
\]
Of course the equilibria of these differential equations are precisely 
the AI states, i.e., any sequence of critical points, $x_{j}= 
c_{k_{j}} \,,\, k_{j} \in\{1,2,\ldots m\}, j \in \Z$.  Such an 
equilibrium is a sink if $s_{j}= sgn(V''(c_{k_{j}}))$.  For example
when the $c_{k}$ are minima of $V$, then $s_{j}=1$ as we expect. 
The basin of attraction of the sink is the box
\[
    D_{\bs}(0) = \{ \bx: c_{k_{j}-1} < x_{j} < c_{k_{j}+1} \}  \,.
\]
Since hyperbolic fixed points of a vector
field are preserved under a $C^{1}$ perturbation \cite{Hirsch74}, we 
immediately have the theorem
\begin{theorem}\label{AIBWtheorem}
   If $L = \epsilon T(x,x') - V(x) \in C^{2}$, and the critical points 
   $\{c_{k}\}$ of $V$ are nondegenerate, then there is an 
   $\epsilon_{max}$ such that for all $|\epsilon| < \epsilon_{max}$ and 
   all sequences of critical points $\{c_{k_{j}}, j \in \Z\}$, the system 
   of differential equations \eq{BWode}, with $s_{j}= 
   sgn(V''(c_{k_{j}}))$, has a hyperbolic sink, $\bx^{*}$, that continues 
   from $x_{j} = c_{k_{j}}$ at $\epsilon = 0$ .
\end{theorem}

This reformulation of the Biham-Wenzel method points out several 
important things.  First, the correct choice of the signs is 
determined by the signature of the potential $V$ at the appropriate 
critical point.  In fact, as originally formulated, the method cannot 
possibly find all orbits when $V$ has more than two critical points.  
Rather, an orbit is determined first by the choice of anti-integrable 
state $c_{k_{j}}$; the sequence of signs $s_{j}$ is determined 
subsequently.  Finally, as we will explicitly demonstrate below, it is 
sensible to use the AI state as the initial condition for the method.

For the \hen map, we use the scaled Lagrangian to obtain the system
\begin{equation}\label{BWscaled}
           \frac{dz_{j}}{d\tau} =  s_{j} \left(
                                            \epsilon(z_{j+1} + bz_{j-1}) -  z_{j}^{2} +1
                                           \right) \,.
\end{equation}
When $\epsilon=0$, the differential equations decouple and give the
simple equations
\[
           \frac{dz_{j}}{d\tau} =  s_{j} \left( 1-z_{j}^{2} \right) \,,
\]
for which there is a unique attracting fixed point at $z_{j}=s_{j}$ 
with  a basin of attraction
\[
        D_{\bs}(0) = \{ \bz: -1 < z_{j} < \infty \} \,.
\]
We can use \Th{existtheorem} to obtain a bound on $\epsilon$ for the 
persistence of the hyperbolic fixed points of \eq{BWscaled}.  We know 
from this theorem that there is a fixed point $\bz^{*}$ for each $\bs$ 
when $\gamma < \gamma_{\infty}$.  The linearization of 
\eq{BWscaled} around this fixed point is
\begin{eqnarray} \label{BWlinear}
           \frac{d\zeta_{i}}{d\tau} &=&  \sum_{j \in \Z} A_{ij} \zeta_{j} \,,\\ \nonumber
           A_{ij} &=& s_{i} \left[ \epsilon(\delta_{i+1,j} + b\delta_{i-1,j}) -  
                                  2 z_{j}^{*}\delta_{i,j}
                          \right]        \,.
\end{eqnarray}
By the Gerschgorin circle theorem, the eigenvalues $\lambda_{k}$ of ${\bf A}$ 
are contained within the union of the disks centered at the diagonal  
elements with radius given by the sum of the magnitudes of the  
off-diagonal elements \cite{Stoer80}.  Since the row sum is bounded by $ 
|\epsilon|(1+|b|) = \gamma$, we obtain
\[
     \lambda_{k} \in \bigcup_{j} \{\lambda :\big|\lambda + 2|z_{j}^{*}|\big| \le  \gamma \} \,.
\]
According to \Th{existtheorem}, $\bz^{*} \in B_{M_{\infty}}(\bs)$, 
where $M_\infty$ is given by \eq{ballsize}; therefore,
\begin{equation}
\label{lambda}
           Re(\lambda) \le -2(1-M_\infty)+ \gamma \,.
\end{equation}
This is negative precisely when $\gamma < \gamma_{\infty}$. Thus 
we have proven

\begin{theorem}
\label{BHtheorem}
        When $|\epsilon|(1+|b|) < \gamma_{\infty}$, then the orbit of the \hen map 
        labeled by the AI symbol sequence $\bs$ is a hyperbolic sink 
        for the system \eq{BWscaled}.
\end{theorem}

The basin of attraction of the sink initially includes the AI state 
$\bz = \bs$. We will next show that this remains true providing 
$\epsilon$ is small enough. Thus the proper initial condition to use for 
the Biham-Wenzel equations is the AI state.

\begin{theorem}\label{BHbasin}
The AI sequence $\bs$ is in the basin of attraction for the fixed point
$\bz^{*}$ of \eq{BWscaled} providing
\[
    |\epsilon|(1+|b|) < 0.555668 \,.
\]
\end{theorem}

\noindent\underline{Proof: }
Suppose $\bz^*$ is an orbit of the \hen map for some fixed $\gamma 
< \gamma_\infty$.  Using \eq{BWscaled}, the deviation $\bdz = 
\bz-\bz^{*}$ from this orbit satisfies the system of equations
\begin{equation} \label{deltaode}
    \frac{d \zeta_{i}}{d\tau}\ = \sum_{j \in \Z} A_{ij}\zeta_{j} - s_{i} \zeta_{i}^{2} \,,
\end{equation}
where ${\bf A}$ is given in \eq{BWlinear}. It is easy to see that
\[  
       \frac{d}{d \tau} ||\bdz||_\infty \le(-2||\bz^*||_\infty + 
              \gamma + ||\zeta||_\infty)||\zeta||_\infty  \,,
\]
so $||\bdz||_\infty$ decreases when 
\[ 
     0 < ||\bdz||_\infty < 2(1-M) - \gamma \,,
\]
This implies that the basin of attraction of the sink contains this ball:
\[
         D_{\bs}(\gamma) \supset \{ \bz: ||\bz-\bz^{*}|| < 2(1-M) - \gamma \}
\]
If the AI state $\bs$ is to be in this basin, then we require that 
$\bs \in D_{\bs}(\gamma)$, but since we know that $||\bz^{*} - 
\bs||_{\infty} \le M_{\infty}$, this is true when
\[ 
     3 M_\infty(\gamma) < 2 - \gamma \Rightarrow \gamma < 0.55566  \,.
\]
\qed

%
\Sec{Conclusions}{Conc}

A large class of dynamical systems can be formulated to have a 
nondegenerate AI limit.  In this case a simple contraction mapping 
argument can be used to show that many of the symbolic orbits at the 
AI limit persist away from the limit.  We have applied these ideas to 
the \hen map to show that all possible bounded orbits exist in the 
range \eq{epsmax}.  Translating this back to the original parameters 
$k$ and $b$ of the map
\eq{henonmap} gives
\[
     k > \frac{5+2\sqrt{5}}{4} (1+|b|)^{2}
\]
which is exactly the same bound as that found by Devaney and Nitecki 
\cite{Devaney79} using a geometrical argument for the existence of a 
Smale horseshoe.  The advantage of the AI argument is that it can be 
easily generalized to systems where the geometrical argument might be 
difficult, such as higher dimensional maps. Moreover, the AI theory 
can also be used to give bounds for the existence of orbits 
corresponding to various subshifts of finite type \cite{Sterling98}.

Our main interest in the AI limit is to use it as the basis for a 
numerical method to find orbits of various types.  One such method we 
discussed here is the ``pseudo-gradient'' algorithm of Biham and Wenzel.  
We showed that this method, which was previously only justified 
heuristically, has a rigorous foundation close enough to an AI limit.  
Indeed, the signs in the matrix $\mbox{Diag}(\bs)$ are 
determined by signature of the critical point at the AI limit.  We 
also were able to show that the AI state can be used as an initial 
condition for the method.  However, our theorem applies only for a 
limited range of parameter values, and one might reasonably apply the 
method over a wider range of parameters.

Some care is advisable however, since it is known that the method of 
Biham and Wenzel can fail far from the AI limit \cite{Grassberger89}.  
We will show in a forthcoming paper \cite{Sterling98} that in some 
applications, such as that in \cite{Davis91}, the method does appear 
to find all orbits.

Rather than use the Biham and Wenzel method we implement a simple 
``continuation'' technique for our numerical studies 
\cite{Sterling98}.  This method is guaranteed to find all bounded 
orbits of the map that extend to the AI limit (there could be 
``bubbles'' that do not extend, but as far as we know these have never 
been seen for the \hen case).  For example, in the area preserving case ($b=1$), our 
numerical results indicate that the horseshoe is destroyed at 
$\epsilon = 0.41887923$, when a pair of orbits homoclinic to the 
fixed point, $\bs = \{\ldots,1,1,\ldots\} \equiv \{+^{\infty}\}$, 
collide in a saddle-node bifurcation.  These orbits have the symbol 
sequences
\[
     \{+^\infty---+^\infty\} \Longleftrightarrow \{+^\infty-+-+^\infty\} \,.
\]
This parameter value also corresponds to the accumulation point of an infinite 
number of periodic saddle-node bifurcations, and hence the first value 
at which the bounded orbits of the \hen map have nonzero measure.

%
\bibliographystyle{unsrt}
\bibliography{Hen}

\begin{thebibliography}{10}

\bibitem{Aubry95}
S.~J. Aubry.
\newblock Anti-integrability in dynamical and variational problems.
\newblock {\em Physica D}, 86:284--296, 1995.

\bibitem{Aubry90}
S.~Aubry and G.~Abramovici.
\newblock Chaotic trajectories in the standard map, the concept of
  anti-integrability.
\newblock {\em Physica D}, 43:199--219, 1990.

\bibitem{MacKay92}
R.S. MacKay and J.D. Meiss.
\newblock Cantori for symplectic maps near the anti-integrable limit.
\newblock {\em Nonlinearity}, 5:149--160, 1992.

\bibitem{Biham89}
O.~Biham and W.~Wenzel.
\newblock Characterization of unstable periodic orbits in chaotic attractors
  and repellers.
\newblock {\em Phys. Rev. Lett.}, 63:819--822, 1989.

\bibitem{Biham90}
O.~Biham and W.~Wenzel.
\newblock Unstable periodic orbits and symbolic dynamics of the complex
  {H}\'enon map.
\newblock {\em Phys. Rev. A}, 42:4639--4646, 1990.

\bibitem{Davis91}
M.J. Davis, R.S. MacKay, and A.~Sannami.
\newblock Markov shifs in the {H}\'enon family.
\newblock {\em Physica D}, 52:171--178, 1991.

\bibitem{Wenzel91}
W.~Wenzel, O.~Biham, and C.~Jayaprakash.
\newblock Periodic orbits in the dissipative standard map.
\newblock {\em Phys. Rev. A}, 42:6550--6557, 1991.

\bibitem{Biham92}
O.~Biham and M.~Kvale.
\newblock Unstable periodic orbits in the stadium billiard.
\newblock {\em Phys. Rev. A}, 46:6334--6339, 1992.

\bibitem{Dey95}
B.~Dey.
\newblock Unstable periodic orbits and characterization of spatial chaos in a
  nonlinear monatomic chain at the {T}=0 first-order phase transition point.
\newblock {\em Phys. Rev. B}, 52:220--224, 1995.

\bibitem{DAlessandro90}
G.~D'Alessandro, P.~Grassberger, S.~Isola, and A.~Politi.
\newblock On the topology of the {H}\'enon map.
\newblock {\em J. Phys. A}, 23:5285--5294, 1990.

\bibitem{Hansen92}
K.T. Hansen.
\newblock Remarks on the symbolic dynamics for the {H}\'enon map.
\newblock {\em Phys. Lett. A}, 165:100--104, 1992.

\bibitem{Politi92}
A.~Politi and A.~Torcini.
\newblock Towards a statistical mechanics of spatiotemporal chaos.
\newblock {\em Phys. Rev. Lett.}, 69:3421--1324, 1992.

\bibitem{Hunt96}
B.~Hunt and E.~Ott.
\newblock Optimal periodic orbits of chaotic systems occur at low period.
\newblock {\em Phys. Rev. E}, 54:328--337, 1996.

\bibitem{Skodje90}
R.~Skodje and M.~Davis.
\newblock Statistical rate theory for transient chemical species: Classical
  lifetimes for periodic orbits.
\newblock {\em Chem. Phys. Lett.}, 175:92--100, 1990.

\bibitem{Fang94}
H.P. Fang.
\newblock Dynamics for a two-dimensional antisymmetric map.
\newblock {\em J. Phys. A}, 27:5187--5200, 1994.

\bibitem{Kaplan93}
H.~Kaplan.
\newblock Type {I} intermittancy for the {H}\'enon family.
\newblock {\em Phys. Rev. E}, 48:1655--1669, 1993.

\bibitem{Nagai97}
Y.~Nagai and Ying-Cheng Lai.
\newblock Characterization of a blowout bifurcation by unstable periodic
  orbits.
\newblock {\em Phys. Rev. E}, 55:R1251--R1254, 1997.

\bibitem{Grassberger89}
P.~Grassberger, H.~Kantz, and U.~Moenig.
\newblock On the symbolic dynamics of the {H}\'enon map.
\newblock {\em J Phys A}, 22(24):5217--5230, 1989.

\bibitem{Devaney79}
R.L. Devaney and Z.~Nitecki.
\newblock Shift automorphisms in the {H}\'enon mapping.
\newblock {\em Commun. Math. Phys.}, 67:137--146, 1979.

\bibitem{Aubry91}
S.~J. Aubry.
\newblock The concept of anti-integrability: Definition, theorems and
  applications to the standard map.
\newblock {\em Twist Mappings and their Applications, Ed Richard McGehee,
  Kenneth R. Meyer}, pages 7--54, 1992.

\bibitem{Hirsch74}
M.W. Hirsch and S.~Smale.
\newblock {\em Differential Equations, Dynamical Systems and Linear Algebra}.
\newblock Academic Press, 1974.

\bibitem{Stoer80}
J.~Stoer and R.~Bulirsch.
\newblock {\em Introduction to Numerical Analysis}.
\newblock Springer-Verlag, 1980.

\bibitem{Sterling98}
D.~Sterling and J.D. Meiss.
\newblock Homoclinic bifurcations for the area-preserving {H}\'enon map.
\newblock {\em In Progress}, 1998.

\end{thebibliography}

\end{document}